# SPATIAL FLUORESCENCE CROSS CORRELATION SPECTROSCOPY BY MEANS OF A SPATIAL LIGHT MODULATOR

*Yoann Blancquaert, Jie Gao, Jacques Derouard and Antoine Delon*[*]

Laboratoire de Spectrométrie Physique
Université Grenoble I - UMR 5588 CNRS
BP 87, 38402, Saint Martin d'Hères Cedex, France

Spatial Fluorescence Cross Correlation Spectroscopy is a rarely investigated version of Fluorescence Correlation Spectroscopy, in which the fluorescence signals from different observation volumes are cross-correlated. In the reported experiments, two observation volumes, typically shifted by a few µm, are produced, with a Spatial Light Modulator and two adjustable pinholes. We illustrated the feasibility and potentiality of this technique by: i) measuring molecular flows, in the range 0.2 – 1.5 µm/ms, of solutions seeded with fluorescent nanobeads or rhodamine molecules (simulating active transport phenomenons); ii) investigating the permeability of the phospholipidic membrane of Giant Unilamellar Vesicles *versus* hydrophilic or hydrophobic molecules (in that case the laser spots were set on both sides of the membrane). Theoretical descriptions are proposed together with a discussion about Fluorescence Correlation Spectroscopy based, alternative methods.

* Corresponding author: adelon@ujf-grenoble.fr, Phone: +(33) 4 76 63 58 01, Fax: (33) 4 76 63 54 95

# 1. Introduction

Since its invention in the early seventies [1], Fluorescence Correlation Spectroscopy (FCS) has been used mostly to measure diffusion coefficients and concentrations. About ten years ago it has been extended to spatial cross-correlation spectroscopy (sFCCS), which is based on a time and spatial dual correlation. It usually implies two shifted laser spots, creating two separate observation volumes from which the fluorescence signals are detected. This method can thus assess the directional transport of molecules or particles between these volumes.

One key point, when performing sFCCS, is the fabrication of separate laser spots. Several techniques have been already used for doing this. In one of the pioneering experiments, based on elastic scattering of light by latex spheres, the two volumes were obtained with two different $Ar^+$ laser lines (458 and 514 nm) [2]. A few years later, two papers reported sFCCS under confocal geometry [3,4]. Interestingly enough, it has been demonstrated that time gated, two photon excitation, significantly lowers the undesired cross-talk between the two observation volumes [4]. However, all these experiments used polarizing beam splitters and/or Wollaston prisms to split and recombine the two, almost parallel, laser beams [3-5]. This made the adjustment of the separation between the two observation volumes not possible. Nevertheless, a Wollaston prism is an interesting optical component because it can be easily inserted in the illumination path of commercial FCS systems, providing a fixed and known distance between the two foci which can therefore be used as an external ruler [5,6]. Passive Diffractive Optical Elements (DOE) have also been used to produce gaussian foci of submicrometer diameter and perform FCS and sFCCS experiments [7,8]. However, here also there is some lack of flexibility, since each DOE is fabricated once and for all by electron beam lithography. A significant improvement has been recently brought to spatial measurements by scanning FCS [9]. This technique has been applied with various modalities [10-12] and can even be extended by exploiting the time structure present in images obtained with laser scanning confocal microscopes [13]. Finally, on the route towards multiconfocal FCS we note the promising approaches using either EM-CCD camera [14], where each pixel is a unitary pixel, or spinning disk confocal microscopes [15], where up to $\sim 10^5$ independent locations are available. Therefore, it appears clearly that the ideal multiconfocal FCS experiment would involve: i) a flexible way to address, simultaneously, the desired laser spots at various locations within the biological medium; ii) a matrix of fast, point-like detectors. Concerning the first aspect, Spatial Light Modulators (SLM) became recently commercially available, opening new possibilities, especially for optical tweezers [16], since these devices make it possible to control the laser illumination geometry.

In this paper we demonstrate the possibility of controlling the illumination with a SLM, by performing sFCCS measurement of active flow and permeability through the membrane of Giant Unilameller Vesicles (GUV). The article is organised as follows: in the following section 2) we present the experimental set-up with its main characteristics, the spatial properties of the laser beam after reflexion on the SLM and the corresponding consequences for FCS data; section 3) is mainly devoted to the theoretical approach of permeability measurement with sFCCS; the experimental results are presented in section 4); finally we conclude in section 5).

# 2. Experimental

## *2.1 Experimental set-up*

In our experimental set-up (Fig. 1), the fibered incident laser beam of power ≅ 200 µW (488 nm, KYMA, Melles Griot), was firstly collimated and expanded (2.4×) through a telescope (lenses 1 and 2), then sent to the SLM device (LC-R 720, Holoeye), by means of mirrors 1 and 2. A $\lambda/2$ plate and a polarizer were introduced before the SLM to optimize the relative intensity of the various spots at the focus plane of the objective. Our SLM device (a reflective LCOS microdisplay) has 1280×768 pixels, a pitch of 20 µm and can modulate the phase (encoded on 8 bits) between 0 and $\pi$. When necessary, the SLM could be bypassed by tilting mirrors 1 and 3.

Then, the laser beam, reflected by a periscope (not shown) and a dichroic mirror was focused within the sample, through the water objective (60×, NA 1.2) of an inverted microscope (Olympus IX70). The fluorescence light was spectrally filtered, separated in two equivalent paths by means of a beam splitter and focused on two multimode fibers (core diameter 100 µm), used as point detectors. Note that the pairs of lenses 3, 4 and 3, 5 produce an additional 3× magnification, such that the total magnification was about 180×. Finally, the two single photon detectors (Avalanche Photodiodes, APD) were connected to a home made data acquisition system and correlator.

We imaged various phase gratings on the SLM: echelette grating, where the phase varies linearly from 0 to $\pi$ over a period of $P_{xl}$ pixels; binary grating, where $\frac{1}{2}P_{xl}$ pixels have no phase shift, followed by $\frac{1}{2}P_{xl}$ pixels with a phase shift of $\pi$ (period of $P_{xl}$ pixels); phase step 0-$\pi$. It is interesting to note that the distance $d$ between the $0^{th}$ and $1^{rst}$ order spots of an echelette grating of period $P_{xl}$ is the same than the distance between the two spots of orders ±1 of a binary grating of period $P_{xl}$ (a binary 0-$\pi$ grating has, theoretically, no $0^{th}$ order spot, but a tiny spot could nevertheless be observed).

The distance $d$ between the centers of the laser spots could be varied between 9.2 µm and 0.8 µm. The longer distance was obtained with a binary grating of period $P_{xl}$ = 20 pixels (corresponding to a spatial period, on the SLM, of 20×20µm = 400 µm) and the shorter one with a 0-π phase step.

For flow measurements, a piezoelectric device (Piezosystem, Jena) has been used, the speed of which lies in the range 200 – 1500 µm/s. In that case a droplet of solution, seeded with a nM concentration of the molecules or particles of interest, was deposited on a cover slit attached to the piezoelectric device.

## *2.2 Spatial properties of the laser beam*

Since it is known that SLM devices introduce geometrical aberrations [17], we have characterized the intensity profile of the laser beam with a CCD camera (Spot Insight B/W, Diagnostic Instruments) and its wave-front with a Shack-Hartmann sensor (HASO32, Imagine Optics). Note that these optical devices, inserted before the entrance of the microscope, are not shown in the Fig. 1 scheme. We first recorded images of the beam profile at various distances from the focal plane of an achromatic doublet (FL300, Melles Griot). Calculating the moments of the intensity distribution [18] and using the standard equations for the gaussian beam propagation [19], we obtained reliable values of the $M^2$ parameters along two perpendicular directions, as shown in Table 1. The $M^2$ factor, also called the beam quality, is the ratio of the beam's divergence to that of a diffraction-limited beam of the same waist diameter [19]. When the laser beam was reflected by the SLM, a blank or flat signal (*i.e.* no spatial variation of the phase shift) was addressed to it.

Table 1 clearly indicates that geometrical aberrations are introduced, among which there must be, at least, astigmatism. Therefore, we measured these aberrations by using the HASO device (Fig. 2). We quantified these aberrations with the Zernike polynomial expansion. Getting rid of the tilt parameters which are not relevant in the present study, we found that the next and dominant, lowest order terms of the Zernike expansion, are the focus and the astigmatism coefficients [20], the values of each of them were of the order of 0.1 µm. The exact values were dependent upon the location of the laser beam on the LCOS micro display and upon the flatness of its surface, which is mechanically deformable. Then, we investigated the possibility of correcting the distorted wave-front by addressing spherical and astigmatism corrections to the SLM [21]. However, the efficiency of this method was found to be quite limited because the amplitude of the phase modulation of our LCOS device is λ/2, while the total wave-front distortion overcomes this value (see Fig. 2).

## *2.3 Fluorescent probes and sample preparation*

Experiments about permeability have been performed with single dye molecules, the lipophilic Rhodamine 6G (R6G) and the hydrophilic Sulforhodamine G (SRG), both from Radiant Dyes. These molecules were used without further purification and diluted at concentrations between 1 and 10 nM. In addition, to study active transport, we also used fluorescent nanobeads of diameter 20 nm (Fluospheres F-8888, Invitrogen).

The Giant Unilameller Vesicles (GUV) used for permeability assessment were swollen from a lipid solution (DOPC) using the electroformation method [22]. They were prepared in sucrose solutions of concentration 270 mM. The GUV were then diluted in glucose solutions (300 mM) of equal osmolalities, in order to sediment on the bottom of the perfusion chamber, where they could be observed. The concentrations of fluorophores (SRG or R6G) inside and outside the GUV, were adjusted so that the inner one matches the outer one.

## *2.4 Properties of the FCS confocal volume*

It appeared interesting to evaluate the influence of the optical aberrations upon the FCS measurements in solution, that is upon the corresponding confocal (or observation) volume. By fitting, with the standard diffusion model, the autocorrelation function of the fluorescence signal [23], we could determine the characteristic diffusion time, $\tau_D = \omega_r^2/4D$ ($\omega_r$ being the radius of the usual gaussian describing the confocal volume and $D$ the diffusion constant). Firstly, when the laser beam bypassed the SLM, the measured diffusion time of R6G in pure water, $\tau_D$, was about 29 µs. This corresponded to $\omega_r$ = 0.18 µm, assuming $D$ = 280 µm$^2$/s for the diffusion constant, at room temperature, of R6G in water [24]. Secondly, when the laser beam was reflected by the SLM (with a blank or flat signal), the diffusion time increased to about 45 µs. This corresponded to an increase of $\omega_r$ to 0.22 µm, due to the aberrations introduced by the SLM reflecting the laser wave-front. In practice, any deformation of the wavefront and, henceforth of the intensity distribution at the focal point of the objective, corresponds to an increase of the effective confocal volume, since it is the product of the illumination function and of the detection one. However, theoretically, one must make a distinction between: i) the focus, that changes the beam cross-section at the focal point and that could be corrected by adjusting the convergence of the incident beam; ii) the astigmatism, that breaks the anisotropy of the confocal volume.

It must also be noted that the viscosity of the sucrose (270 mM) and glucose (300 mM) solutions used for the permeability experiments, is significantly different from that of pure water [25]. As a first consequence, the corresponding diffusion time increased to 52 µs for glucose ($\eta = 1.16\times10^{-3}$ Pa.s) and 65 µs for sucrose ($\eta = 1.29\times10^{-3}$ Pa.s). In addition, the change of the solution refraction index, pure water *versus* sugar solutions, might have also corrupted the value of $\tau_D$, since the refraction index modifies the laser focalization and fluorescence collection and, in turn, the radius, $\omega_r$, of the observation volume.

From a practical point of view, the enlargement of the confocal volumes increases their overlap. However, despite this phenomenon, sFCCS could be used to access permeability and to measure active transport, providing the observation volumes are farther apart than ≅ 1 µm.

## 3. Theoretical background about sFCCS

Lets us consider the general definition of the sFCCS function, obtained by cross-correlating the fluorescence signals, $I_1(t)$ and $I_2(t)$, emanating from two different observation volumes, spatially shifted by $\mathbf{\Delta}$ :

$$G_{CC}(\mathbf{\Delta},\tau)=\frac{\langle I_1(t)I_2(t+\tau)\rangle}{\langle I_1(t)\rangle\langle I_2(t)\rangle}=1+\frac{\langle \delta I_1(t)\delta I_2(t+\tau)\rangle}{\langle I_1(t)\rangle\langle I_2(t)\rangle} \quad (1)$$

where $\delta I_{1,2}(t) = I_{1,2}(t) - \langle I_{1,2}(t)\rangle$ are the deviations from the temporal averages of the signals. These deviations arise from the fluctuations of the concentration of fluorescent molecules, $\delta C$, within the observation volumes, $W_1$ and $W_2$:

$$\delta I_{1,2}(t)=A_{1,2}\int W_{1,2}(\mathbf{r})\delta C(\mathbf{r},t)d\mathbf{r} \quad (2)$$

where the constant $A_{1,2}$ takes into account the laser intensity, the fluorescence cross-section and the overall fluorescence detection efficiency. The observation volumes are themselves the product of an illumination function and of the detection one [26]. Without entering within the details, we stress the fact that the illumination function is controlled by the laser excitation shaped by means of the SLM, while the detection one corresponds to the pinhole adjustment.

To describe our experimental situation, we consider two shifted, but equivalent, observation volumes, described by the same volume $W$ and the same constant $A$. Therefore, the cross-correlation function reads:

$$G_{CC}(\mathbf{\Delta},\tau)=1+\frac{\iint W(\mathbf{r'}+\mathbf{\Delta}/2)W(\mathbf{r''}-\mathbf{\Delta}/2)\langle\delta C(\mathbf{r'},t)\delta C(\mathbf{r''},t+\tau)\rangle d\mathbf{r'}d\mathbf{r''}}{\langle C\rangle^2\left[\int W(\mathbf{r'})d\mathbf{r'}\right]^2} \quad (3)$$

where $\langle C\rangle$ is the mean value of the concentration.

### 3.1 sFCCS with flow and diffusion

The two laser spots being on a line parallel to the molecular flow of speed $V$, the cross correlation function is found to be, in presence of diffusion [3]:

$$G_{CC}(\Delta,\tau)=1+\frac{1}{N}\exp\left[-\left(\frac{\Delta}{\omega_r}\right)^2\frac{(1-\tau/\tau_V)^2}{(1+\tau/\tau_D)}\right]\frac{1}{(1+\tau/\tau_D)\sqrt{1+\tau/S^2\tau_D}} \quad (4)$$

where: $\omega_r$ and $\tau_D$ are the same parameters as those defined in section 2.4; $\omega_z$ is the elongation of the observation volume; $S = \omega_z/\omega_r$ is the so called structure parameter; $N = \langle C\rangle \pi^{3/2}\omega_r^2\, \omega_z$ is the number of molecules in the effective observation volume; $\tau_V = \Delta/V$ is the characteristic flow time between the two observation volumes.

### 3.2 sFCCS with permeability and diffusion

In presence of permeability, the propagator, that appears within the numerator of the above Eq. 3, contains two contributions. The first one, along the $x$ direction, stands both for the diffusion, with constant $D$ and for the permeability across the phospholipidic membrane, with constant $P$ [27]; the second one stands for the diffusion parallel to the $y$, $z$ plane [28]:

$$\langle\delta C(\mathbf{r'},t)\delta C(\mathbf{r''},t+\tau)\rangle=\langle C\rangle\frac{P}{D}\left\{\exp\left[\frac{2P}{D}(2P\tau+x''-x')\right]erfc\left(\frac{4P\tau+x''-x'}{2\sqrt{D\tau}}\right)\right\}\left\{\frac{\exp\left[-\frac{(\mathbf{r}_{yz}''-\mathbf{r}_{yz}')^2}{4D\tau}\right]}{4\pi D\tau}\right\} \quad (5)$$

where the subscripts $yz$ correspond to vectorial components in the y, z plane of the membrane (located at x = 0). Using ref. [29], one can check that when the permeability $P$ goes to infinity, the corresponding one-dimensional contribution recovers the usual free diffusion expression. Inserting Eq. (5) within Eq. (3) and assuming (as usual) a gaussian shape for $W$ does not lead to an analytical expression of the cross-correlation function. Although the corresponding numerical solution could be fit with a multi-parameter analytical expression, the latter would be painful to manipulate and discuss, so that we prefer to represent $W$ by a delta function to obtain a simple analytical solution:

$$W(\vec{r})=e^{-2x^2/\omega_r^2}\times e^{-2y^2/\omega_r^2}\times e^{-2z^2/\omega_z^2}\rightarrow\left(\frac{\pi}{2}\right)^{3/2}\omega_r^2\omega_z\delta(x)\delta(r_{yz}) \quad (6)$$

Note that the integration of $W$ still leads to a finite observation volume containing the same number of molecules as above mentioned, *i.e.* $N = \langle C \rangle \pi^{3/2} \omega_r^2\ \omega_z$. Using this approximation for the observation volumes (symmetrically shifted by $\pm\Delta/2$ on both sides of the membrane) and inserting it in the cross-correlation expression (Eq. 3), together with the propagator (Eq. 5), led to the following result:

$$G_{CC}(\Delta,\tau)=1+\frac{1}{N}\sqrt{\pi}\omega_r\ \frac{P}{D}\left\{\exp\left[\frac{2P}{D}(2P\tau+\Delta)\right]erfc\left(\frac{4P\tau+\Delta}{2\sqrt{D\tau}}\right)\right\}\left\{\frac{1}{\sqrt{\tau/\tau_D}\sqrt{\tau/S^2\tau_D}}\right\} \quad (7)$$

Once again, this expression contains two contributions, one for the permeability and one for the diffusion. It is interesting to compare the expression of the diffusion contribution (last factor of Eq. 7) with the exact expression for two-dimensional diffusion, that involves the term $(1 + \tau/\tau_D)^{-1/2}(1 + \tau/S^2\tau_D)^{-1/2}$. One clearly sees that the approximation made for the observation volumes (Eq. 6) is correct as long as the observation time, $\tau$, is much larger than the diffusion time through the observation volume, $\tau_D$. We will see in the experimental section that the observed features fulfil this condition, thus justifying the corresponding approximation. It is worthwhile to make a practical remark about the utilization of the Eq. 7: since the argument of the error function is always positive, the approximation by excess presented in ref. [29] is very useful to evaluate the cross correlation function, whatever are the parameters and variable values, $P$, $D$, $\Delta$ and $\tau$.

As a result, we present in Fig. 3 a series of cross-correlation functions calculated with parameters consistent with the experimental values: the permeability, $P$, is ranging from 10 to $10^3$ µm/s (solid curves); $D$ = 280 µm$^2$/s; $\Delta$ = 1.4 µm; $\omega_r$ = 0.22 µm and $S$ = 5. In addition, the limiting case of infinite permeability (dashed curves) is also shown. The first remark concerns the overall amplitude of these curves. Despite the fact that they are calculated for a density that corresponds to an average of $N$=1 molecule in the observation volume (thus corresponding to an auto-correlation function of amplitude 1), their amplitudes are much weaker than 1. This is because the molecules diffuse in a three-dimensional space, implying that only a small fraction of them pass from one observation volume to the other. The second and also disappointing remark is that the location, in time, of the maximum of the correlation functions is weakly dependent upon the permeability, thus preventing one to easily estimate the permeability by measuring this time. The last, but more salient, feature concerns the evolution of the amplitude of the cross-correlation function *versus* the permeability. One observes that when $P$ = 1000 µm/s, the amplitude of the cross-correlation function is close to that of the asymptotic value ($P = \infty$, *i.e.* free diffusion), while, when $P$ = 10 µm/s, its amplitude is negligible. The transition, as a function of $P$, towards the limiting case of free diffusion is governed by the value of the ratio $D/\Delta$: consequently, in our case, when $P$ is much lower than 200 µm/s, the amplitude of the cross-correlation function is very weak; conversely when $P$ is much larger than 200 µm/s the difference with the free diffusion case becomes small.

## 4. Results and discussion

### 4.1 Flow measurements

As presented in the Experimental section, we addressed various kinds of phase modulations to the SLM. The 0-π phase step is the phase pattern that produces the shortest distance between the two laser spots (0.76 µm, measured with a calibrated CCD camera). As comparison, the larger separation was obtained with a binary grating of period $P_{xl}$ = 20 pixels, that produces two laser spots shifted by 9.2 µm. Varying the speed of the flow, $V$, from 0.5 to 1.5 µm/ms, we could easily observed cross correlation functions $G_{CC}(\Delta,\tau)$, the maxima of which being shifted towards shorter times when the speed, $V$, increases (Fig. 4). Typically, 10 acquisitions of 30 s have been recorded, in order to provide averaged values and error bars.

Since the goal of the present work is to demonstrate that SLM makes it possible to produce laser spots (and henceforth observation volumes) of tuneable separations and to make use of it for sFCCS experiments, we have simply characterised the cross-correlation functions by the time of their maxima, $\tau_{max}$ and did not fit the experimental data to get the transport time $\tau_V$ (Eq. 4). One observes, in the insert of Fig. 4, a linear relation between $\tau_{max}$, and $1/V$, from which one can estimate the separation between the two observation volumes, $\Delta$ = 0.94 µm for the 0-π phase step and 8.5 µm for the binary grating of period $P_{xl}$ = 20 pixels. The difference between these distances and those measured between the corresponding laser spots with the CCD camera (0.76 and 9.2 µm) is partly due to the fact that, when the two spots are closed, the two pinholes do not match exactly the two laser spots, in order to limit the optical cross-talk between the two channels. This cross-talk, sometimes called pseudo autocorrelation [4], appears as a back-ground signal at times shorter than the maxima of the cross-correlation functions. As a consequence, the theoretical cross-correlation function given by Eq. 4 does not match properly the observed data when the two spots are closed to each other (data not shown), so that, to obtain a good fit, it would be necessary to introduce an autocorrelation contribution of adjustable weight and appropriate effective autocorrelation time [30]. This is the reason why it was simpler to characterise the cross-correlation functions by the time of their maxima, $\tau_{max}$.

Cross-correlation data similar to those of Fig. 4 have been obtained using a echelette grating. The practical dif-ference being that the two pinholes must, in that case, fit the $0^{th}$ and $1^{rst}$ order laser spots, rather than the laser spots of orders ±1.

We also used molecular probes (R6G) to perform cross-correlations (data not shown). The important feature is that, compared to fluorescent nanobeads, the amplitude of the signal was much weaker because, simultaneously to the active transport of the molecules between the two volumes, diffusion took place much faster. For instance, using the Stokes-Einstein relation, one can estimate the diffusion constant of the 20 nm beads to be about 20 $\mu m^2/s$ at room temperature, that is about 15 times slower than that of R6G (280 $\mu m^2/s$ [24]). Consequently, the diffusion time, $\tau_D$, of beads through the observation volume is about 500 µs (*versus* about 30 µs for R6G), that is of the order of the transport time, $\tau_V$, between the two volumes separated by a few µm, with a flow speed of about 1 µm/ms. This situation is favourable for the observation of nanobeads flow, contrarily to the case of fast diffusing - small molecules such as R6G: as a matter of fact, Eq. 4 immediately indicates that the amplitude of the cross-correlation function, at time $\tau = \tau_V$ (which is close to the time of the maximum, $\tau_{max}$), varies as ~ $(\tau_D/\tau_V)^{3/2}$.

## 4.2 Permeability assessment

We have assessed the permeability of GUV membranes *versus* the hydrophilic or lipophilic nature of the molecules, by setting one of the spots within the GUV and the other one outside. These experiments have been performed using a echelette grating of period 60 pixels, optimised to produce two spots of order 0 and 1, shifted by $\Delta$ = 1.35±0.10 µm (other orders were almost invisible). This can be seen in Fig. 5, together with two GUV.

In order to provide averaged values and error bars 10 acquisitions of 30 s have been recorded. Note that in all cases (see Fig. 6) we observed a relatively large value of the cross-correlation function at short times (< $10^2$ µs), also called pseudo autocorrelation, once again due to the partial overlap between the two observation volumes [4]. This overlap depended drastically upon the distance between the two volumes. In our experiment, their separation resulted from a compromise: if the separation is smaller, the pseudo autocorrelation becomes too large; conversely, if the separation is larger, not enough molecules diffuse from one volume to the other and the amplitude of the corresponding cross-correlation bump becomes too small. Note that the pulsed laser technique, that is sometimes used to significantly decrease the pseudo autocorrelation [4,6], cannot be combined with a SLM, since in this kind of experiment a unique incident laser beam is diffracted by the SLM to produce the two observation volumes.

We first considered the case of hydrophilic molecules, SRG, that do not pass through a lipidic bilayer. We clearly see in Fig. 6 that, while a bump is clearly observable when both spots are set within the glucose solution, this correlation bump disappears as soon as the two spots are set on each side of the membrane. This bump corresponds to the time it takes for the molecules to diffuse from one observation volume to the other. Note that this time (≅ $10^3$ µs) is much larger that the diffusion time through one observation volume ($\tau_D \cong 50 - 60$ µs) thus validating the approximation made in the theoretical section to calculate the cross correlation function ($\tau \gg \tau_D$). Taking into account the overlap between the two observation volumes as a pseudo autocorrelation contribution of characteristic time $\tau_{ac}$ and weight $p$, we performed a fit of the sFCCS function, with $D$ = 241 $\mu m^2/s$ (instead of 280 $\mu m^2/s$, because of the glucose viscosity), $\omega_r$ = 0.22 µm, $S$ = 5 and assuming a free diffusion in the three directions of the space [30]:

$$G_{CC}(\Delta,\tau)=1+\frac{1}{N}\left[p\left(1+\frac{\tau}{\tau_{ac}}\right)^{-3/2}+(1-p)\exp\left[-\frac{\Delta^2}{{\omega_r}^2}\left(1+\frac{\tau}{\tau_D}\right)^{-1}\right]\left(1+\frac{\tau}{\tau_D}\right)^{-1}\left(1+\frac{\tau}{S^2\tau_D}\right)^{-1/2}\right] \tag{8}$$

One can show that the cross correlation contribution of this formula corresponds to the infinite permeability limit of Eq. 7. As can be seen in Fig. 6, this fit (solid curve) provides a reasonable value of the distance $\Delta$ between the two volumes: 1.16 ± 0.02 µm, compared to the measured spot separation (1.35 ± 0.10 µm). These two quantities are not expected to be the same, since the observation volumes result from the combination of the laser illumination (*i.e.* the spot location) and the pinhole detection. We believe that the overlap between the two volumes and the approximate expression, Eq. (8), used to describe the cross correlation also contribute to the difference. Let us now turn to lipophilic molecules, like R6G. Contrarily to SRG, lipidic bilayers are very permeable to these molecules, as shown in the insert of Fig. 6, where one does not observe any significant difference between the cross-correlation curves recorded with, either the two spots embedded within the solution, or set across the membrane. This is because the probability to find lipidic molecules within the membrane is high, as can be easily checked with wide field fluorescence imaging (data not shown). As a consequence, these molecules very quickly pass the membrane. This can be rationalized with the following equation:

$$P = KD_L/\Delta x \tag{9}$$

that relates the permeability, $P$, to the partition coefficient, $K$ and to the diffusion constant, $D_L$, of the molecules of interest in the lipidic bilayer of thickness $\Delta x$ [31]. The (lipids-water) partition coefficient of a given species is the ratio of the concentrations of this species in lipids and in water, at equilibrium. Assuming that the diffusion constants, $D_L$, of R6G and SRG are close together, the difference of their permeability is mainly dependent upon their partition coefficients, *K*. As a matter of fact, while *K* is close to $10^4$ for R6G [32], that of SRG is smaller than 0.01 [33], that corresponds to a much higher solubility of R6G in the lipidic phase. Using the Stokes-Einstein equation and a viscosity η within the phospholipidic bilayer of about 0.01 Pa.s [34], the diffusion constant, $D_L$, is set to 30 µm$^2$/s. Assuming that the thickness, $\Delta x$, of the phospholipidic bilayer is 3 nm, we derived $P$ = $10^8$ µm/s for R6G and $P$ < 100 µm/s for SRG. As discussed in the theoretical section, the cross-correlation bump cannot be observed if the permeability if much lower than 200 µm/s, which is consistent with our observations for SRG (see the Fig. 6). Conversely, the permeability of the membrane to R6G is so large that it does not modify the cross-correlation function, compared to its bulk value, as can be seen in the insert of Fig. 6.

## 5. Conclusion

In this paper we have shown that a SLM device can be used in sFCCS experiments to measure active or passive transport. The principle consists in cross-correlating the intensity fluctuations coming from two nearby observation volumes, the separation of which being controllable by varying the parameters of the computer-addressed phase grating.

Thanks to sFCCS, directional transport can be measured much more easily than can be done with FCS [35]. The technique presented in this manuscript could be applied to flow velocity measurements in micro-scale miniaturized structures in the fields of chemical analysis and biological sciences. Other important applications are the measurements of active transport in living cells, for instance along internal tubular network [36] and the measurements of flow profiles in living tissues and organisms [37]. We also want to stress the fact that our technique could also be used for multi-point FCS measurements, *i.e.* without cross-correlating the photons streams emanating from the different laser volumes. A living cell being a highly inhomogeneous medium, the possibility to assess, by FCS, concentration and diffusion at different places, but simultaneously, is very promising.

Despite the aberrations introduced by the SLM in the optical path, we could easily measure flow speeds of a few µm/ms or less, providing the diffusion time (within each of the confocal volumes) is not too small compared to the transport time between these volumes. The advantage brought by the SLM device is the possibility to optimise the distance between the volumes and to re-orient them, according to the specific geometry of the sample. Our approach can be compared with the one based on scanning FCS and recently implemented on a Laser Scanning Confocal Microscope [12]. The shorter distance we could achieve (0.94 µm) was limited by the overlap between the confocal volumes. However, shorter separations should be attainable if aberration corrections are made (which was not possible with our device, limited to 0-π phase shifts).

Because the permeability is an important property of cellular membrane, that governs exchanges between various domains, we also tested our instrument with GUV, the permeability of which varying with the more or less hydrophobic character of the molecules. Although our results are rather on a yes or not basis, we showed, for the first time, at a single molecule level, the influence of the permeability upon the sFCCS data, by comparing hydrophobic molecules (R6G) that do quickly pass a lipidic membrane and hydrophilic ones (SRG), that do not. One may think about applying the sFCCS technique to cell membranes, such as the nuclear envelope[38] or the membrane connecting neighbouring cells [39]. However, the average permeability of the membranes, that depends upon the area density of transporters (or pores) spanning the membrane (~50 pores / µm$^2$), the structure of the pores (~10 nm in diameter and 40 nm long) and the size of the molecular permeant, is probably too low (< 100 µm/s) to be assessable by sFCCS [see *e.g.* 38-40]. In contrast, the nuclear envelope breakdown occurring during cell division [38] is a situation where the sFCCS technique should be appropriate to measure the density and permeability of disassembled pores.

The two presented experiments are to be considered as proofs of principle that sFCCS is doable with a SLM. Clearly the justification of such a technique comes from live cell studies, where it is highly desirable to perform multiplexed acquisitions, in order to measure, at the same time, molecular concentrations and transport properties at various locations within the cellular environment. An alternative and well established technique would be scanning FCS, but it is limited to the study of relatively slow processes when a large portion of the field of view has to be characterised [9-11]. For this reason, the potentialities offered by EM-CCD cameras, as arrays of pixels, is especially interesting for sFCCS [14], even if, at the moment, these devices are not able to reach high acquisition speeds (to our knowledge, the maximum rate is about 500 images / s for a 128×128 pixel area). Four years ago, it has been shown that a 2×2 array of CMOS-Single Photon Avalanche Detector (SPAD) can be combined with a DOE to perform multi-confocal FCS [41]. Therefore, the emergence, in a probably close future, of CMOS-SPAD cameras with larger number of pixels is very exciting. By combining such a detector with an illumination controlled by SLM, there is, virtually, no limit for sFCCS.

## 6. Acknowledgement

We thank J.P. Pique for the loan of the Shack-Hartmann sensor.

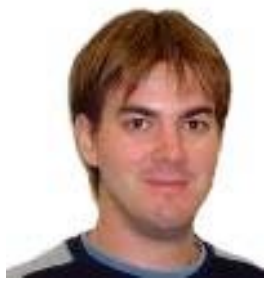

**Yoann Blancquaert**, born 3 August, 1979 in Brest, France. Education: M. Sc in Physical Engineering, Université Grenoble I, France. 2006 PhD in Physical Engineering, Université Grenoble I, France. Current position in a microelectronic company.

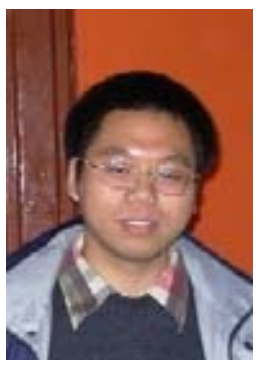

**Jie GAO**, born 21 December, 1981 in Wuhan, China. Education: M. Sc in Physical Engineering, Université Grenoble I, France. PhD student in Laboratoire de Spectrométrie Physique from december 2007, Université Grenoble I, France. Current research interests: Fluorescence Correlation Spectroscopy, cellular biomechanics, microfabrication of polymer structures.

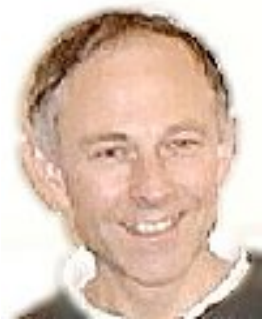

**Jacques DEROUARD**, born 19 June, 1953, is professor at Université Grenoble I, France. Education: 1974 graduated from Ecole Normale Supérieure and University Paris VI. 1983 "Doctorat d'Etat es Sciences", Université Grenoble I, France. Current research interests: optical spectroscopic methods, biomedical optics, single molecule imaging. He has been awarded from CNRS (bronze medal, 1983), and was member of the Institut Universitaire de France (1993-1998)

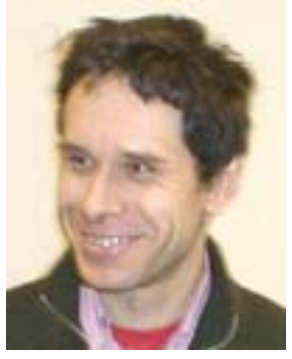

**Antoine DELON**, born 8 December, 1963, is professor at Université Grenoble I, France. Education: 1991 PhD Université Grenoble I, France. Current research interests: Fluorescence Correlation Spectroscopy, single molecule imaging, continuous photobleaching in living cells. He was member of the Institut Universitaire de France (1997- 2002).

# Table and figures

**Table 1** $M^2$ measurements along horizontal (x) and vertical (y) directions.

| Laser beam | $M^2(x)$ | $M^2(y)$ |
|---|---|---|
| Non reflected by the SLM | 1.03 | 1.08 |
| Reflected by the SLM | 2.33 | 5.28 |

------------------------------------------------------------------------------

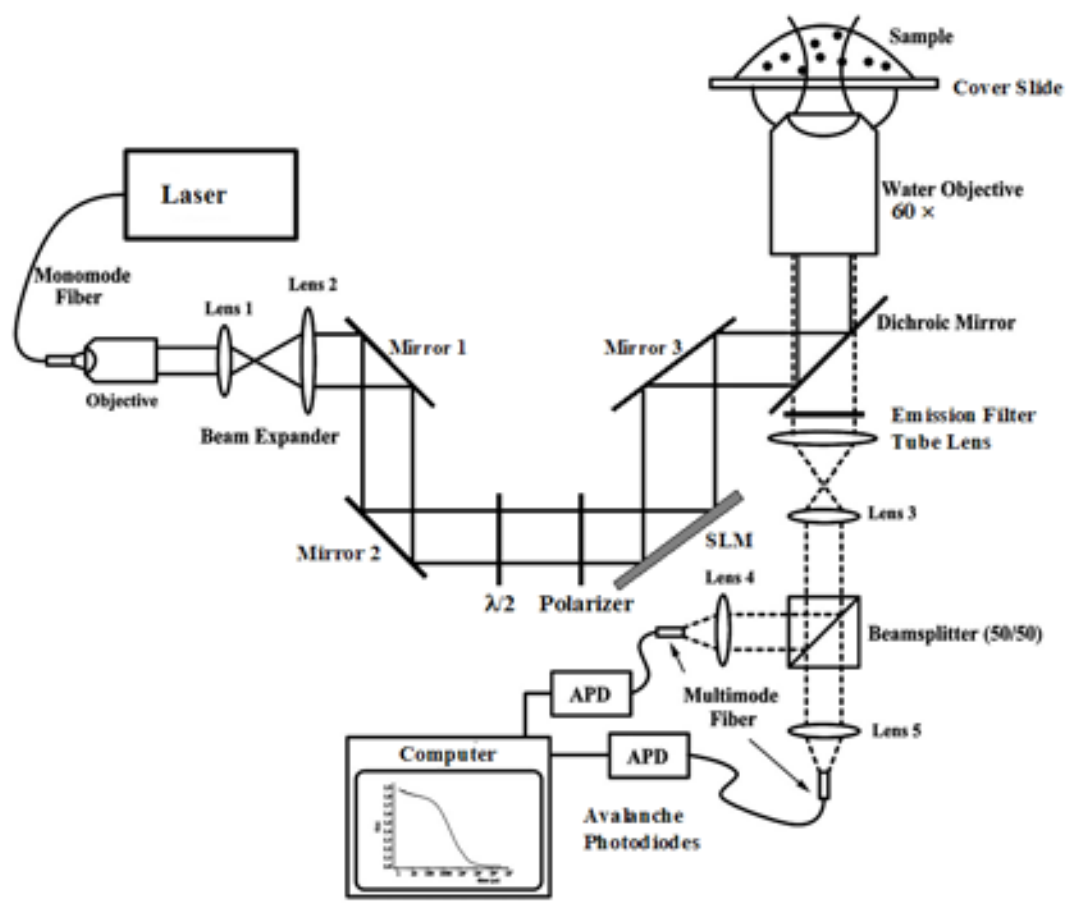

**Figure 1** Experimental set-up, showing the excitation path with the laser shaping optics (beam expander, polarizing elements and SLM device) and the fluorescence path with the spectral filtering, additional magnification and the two independent channels with their adjustable pinholes (optical fibers).

------------------------------------------------------------------------------

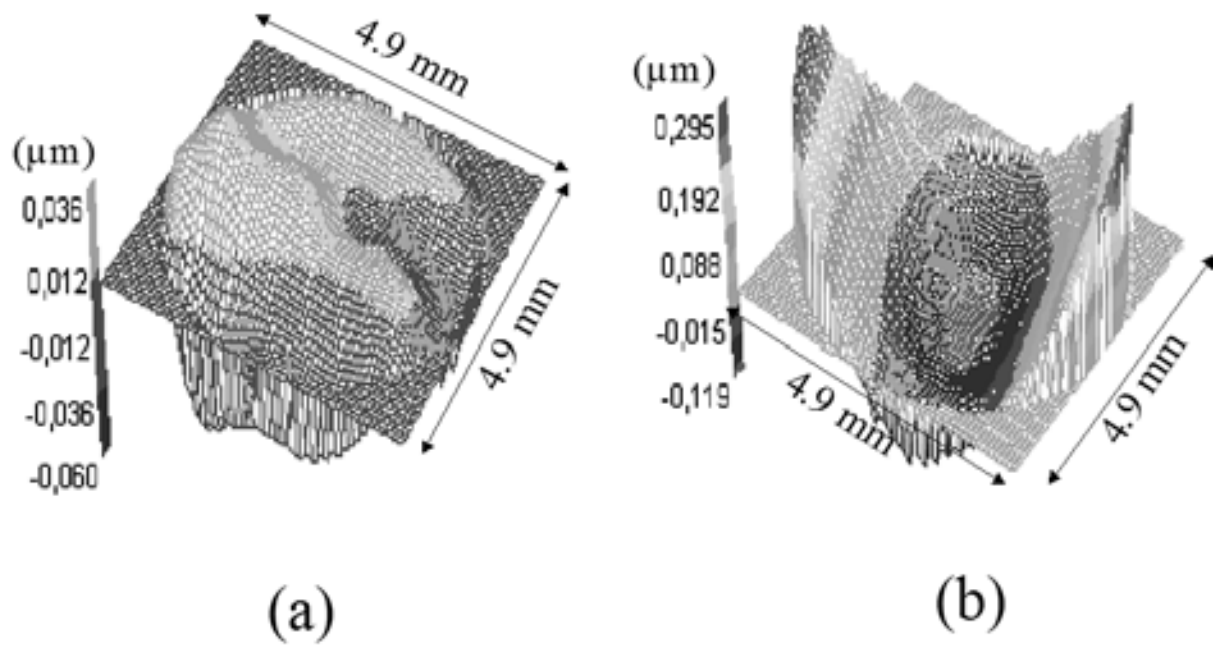

**Figure 2** Laser wavefront, incident (a) or reflected by the SLM (b). Note that the amplitude, peak to peak, of the distortion introduced by the SLM (b) is equal to 0.405 µm (0.295 + 0.119), which is larger than $\lambda/2$ = 0.244 µm. This amplitude (0.405 µm) is also much larger than that of the distortion of the incident beam (a).

------------------------------------------------------------------------------

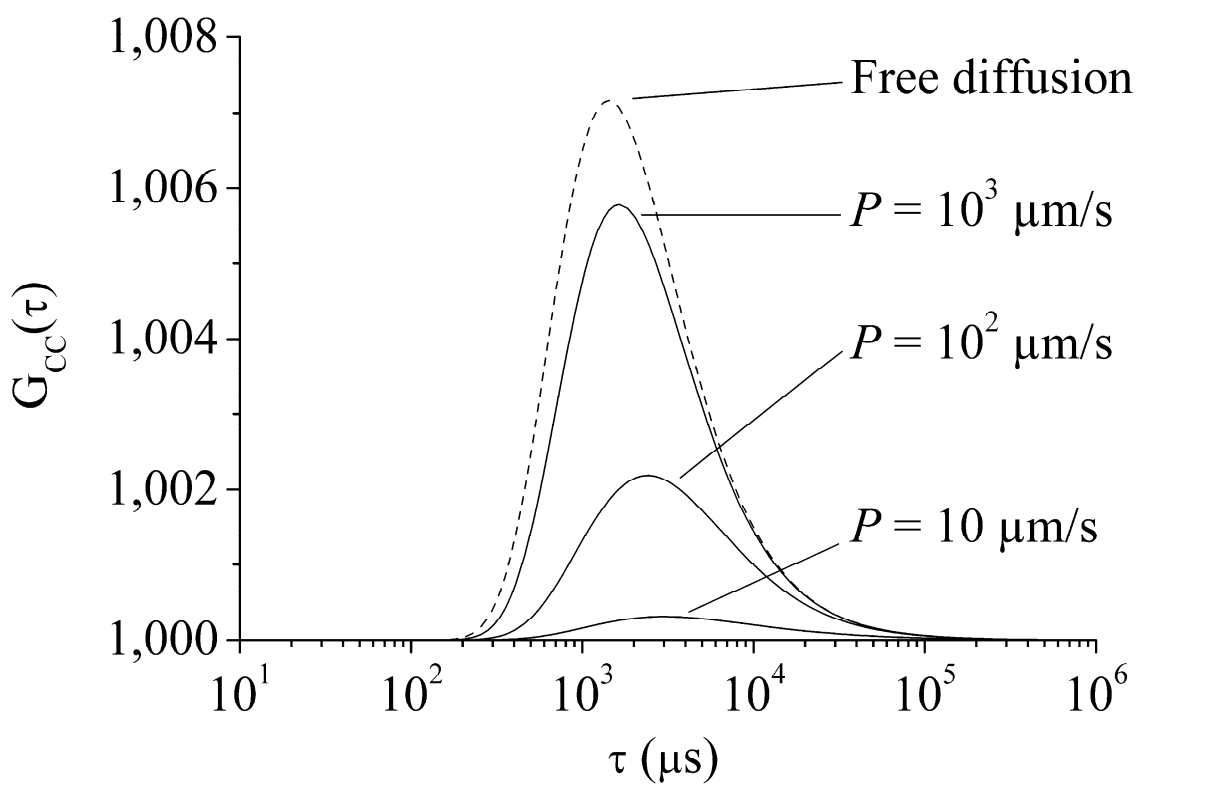

**Figure 3** Calculated cross-correlation functions between two observation volumes shifted by 1.4 µm, assuming a diffusion coefficient of 280 $\mu m^2/s$ and an observation volume radius $\omega_r$ = 0.22 µm. The permeability constant, *P*, is set to 10, $10^2$ and $10^3$ µm/s, in addition to the free diffusion case, *i.e.* *P* = ∞.

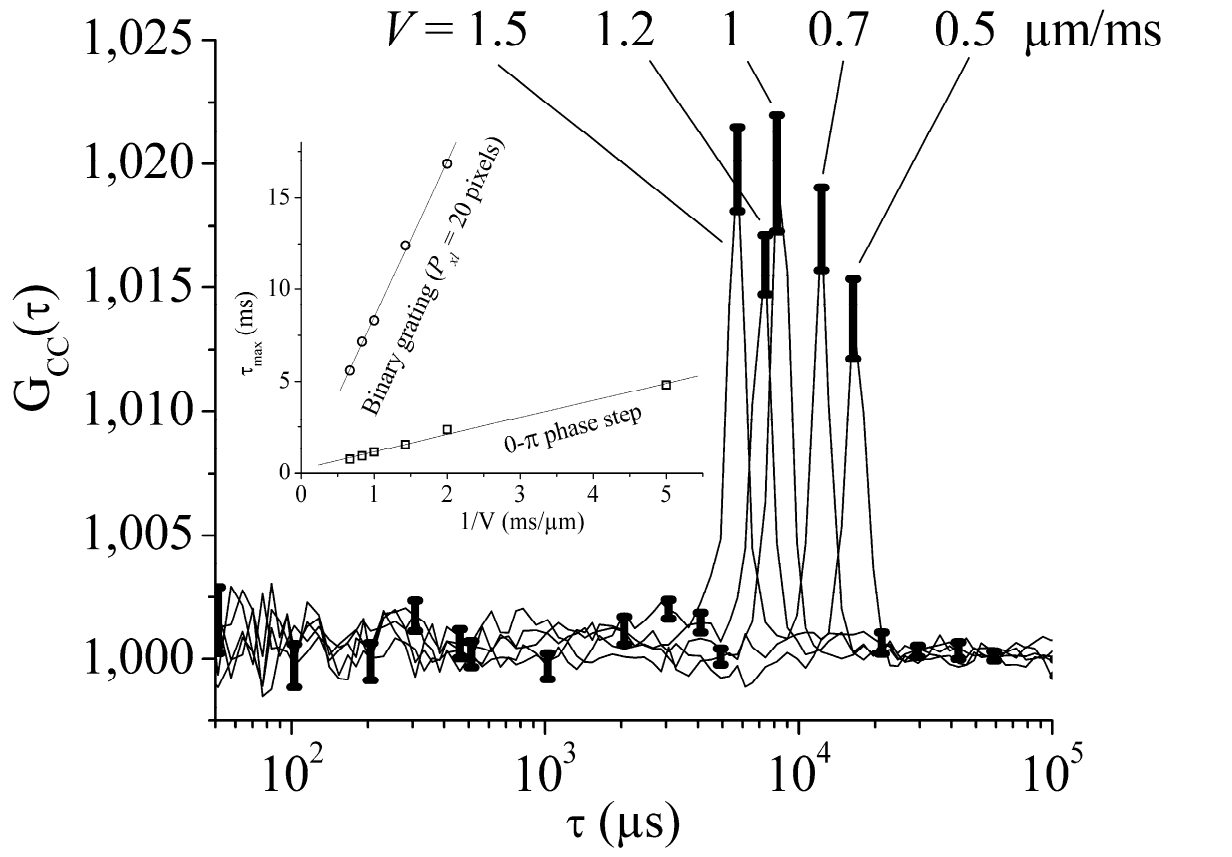

**Figure 4** Cross correlation functions obtained with two observation volumes produced with a binary grating of period $P_{xl}$ = 20 pixels on the spatial light modulator. The fluorescent entities are 20 nm beads and the velocity of the flow is 1.5, 1.2, 1, 0.7 and 0.5 µm/ms. Error bars are standard errors of the mean. The insert shows the times, $\tau_{max}$, of the maxima of the cross correlation functions, *versus* the inverse of the flow speed *V*, for the binary grating and for a 0-π phase step (cross-correlation data not shown). The effective distances between the observation volumes were deduced from linear regressions (straight lines): 0.94 µm for the 0-π phase step and 8.5 µm for the binary grating.

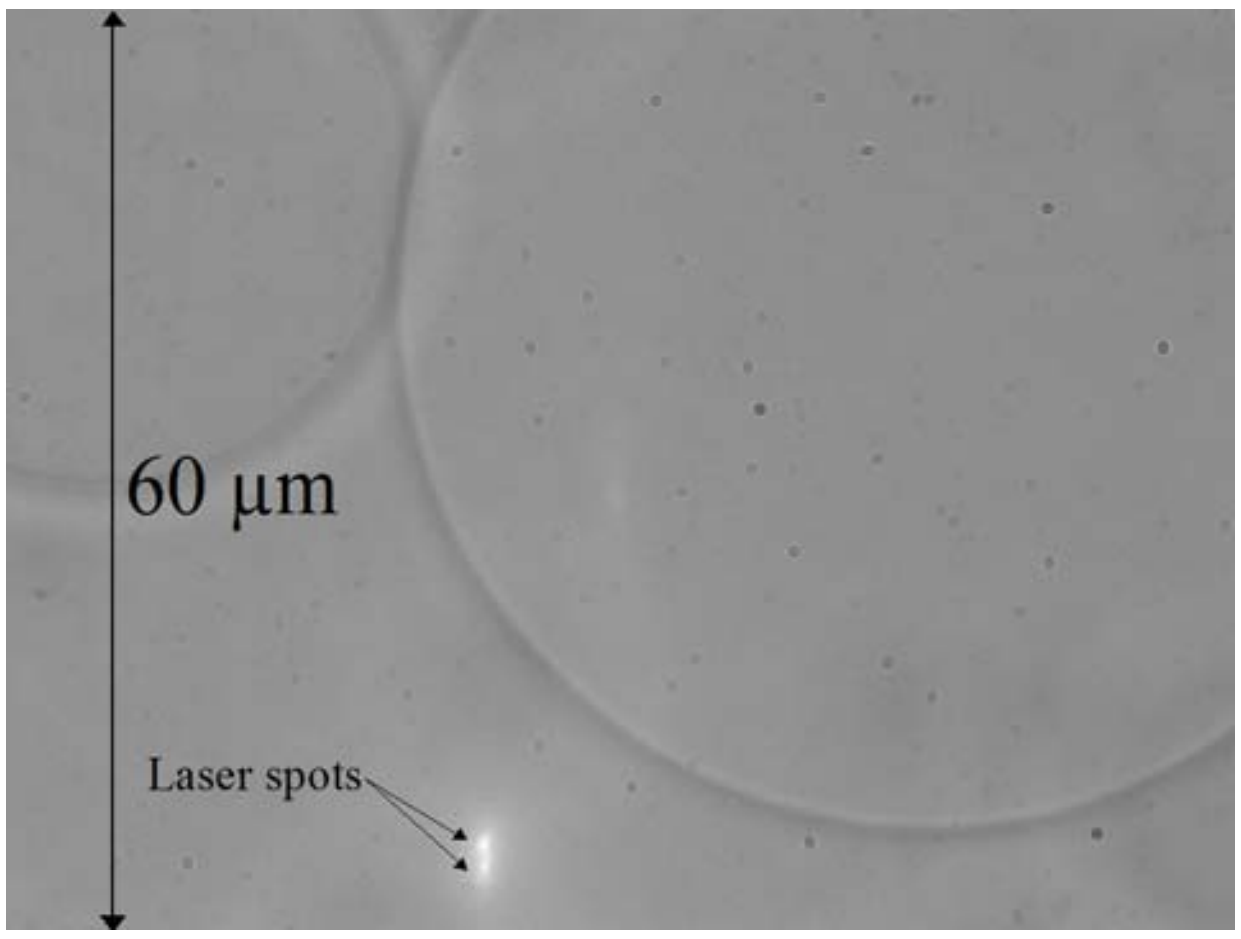

**Figure 5** Wide field view of two GUVs, a few tens of µm in diameter, together with a pair of spots created by the SLM. These two spots, far apart by about Δ = 1.4 µm, can be oriented and placed, either within the inner or outer solution, or across the membrane.

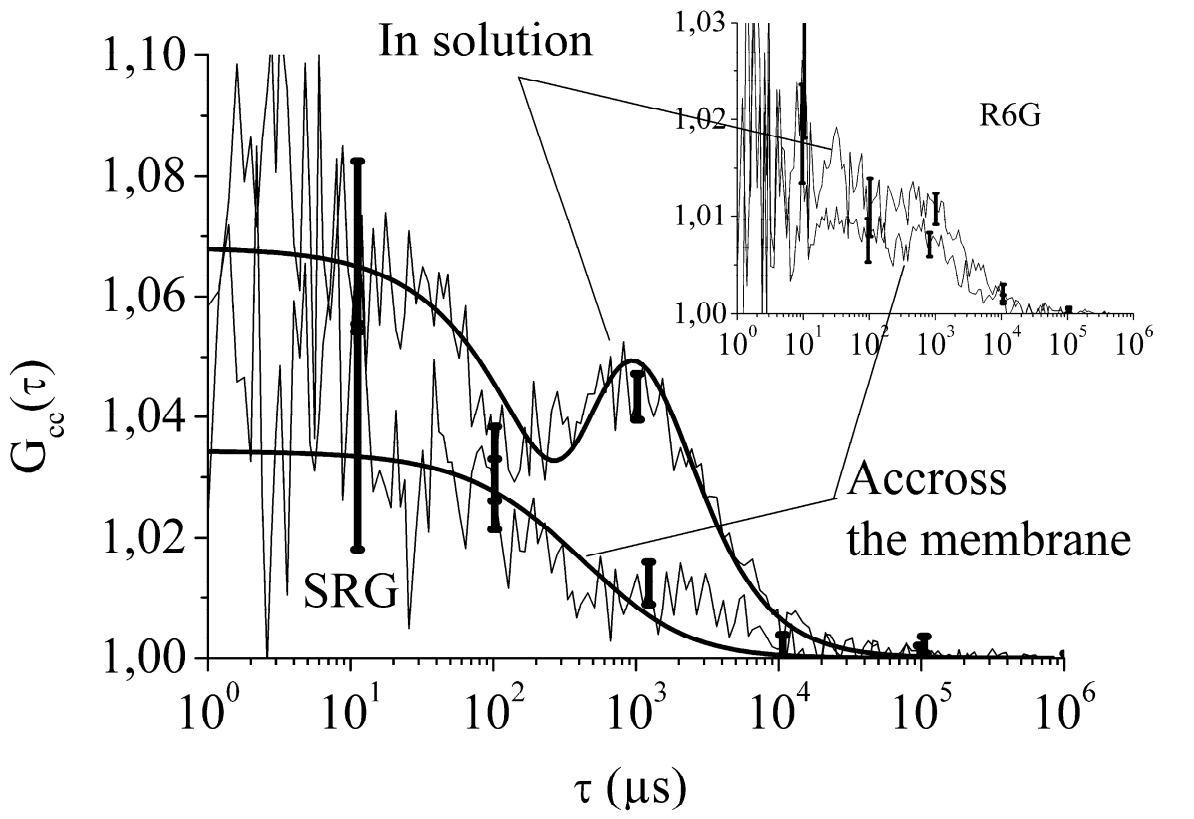

**Figure 6** Cross-correlation functions obtained with SRG, by setting the two spots, either in the external glucose solution, or across the membrane of a GUV. The solid lines are the fits of the cross-correlation functions obtained in solution and across the membrane, taking into account the pseudo auto-correlation and the free diffusion between the two observation volumes. Error bars are standard errors of the mean. The insert shows, for comparison, the cross-correlation functions obtained with R6G; the two curves are closer to each others because of the very large permeability of the membrane for R6G.